\documentclass[12pt]{article}
\input{psfig.sty}

\def\NS{{\cal N}}
\def\N{{ N_c}}
\newcommand{\be}{\begin{equation}}
\newcommand{\ee}{\end{equation}}
\newcommand{\eel}[1]{\label{#1}\end{equation}}
\newcommand{\bea}{\begin{eqnarray}}
\newcommand{\eea}{\end{eqnarray}}
\newcommand{\eeal}[1]{\label{#1}\end{eqnarray}}
\begin{document}

\begin{titlepage}
\rightline{\vbox{\ialign{#\unskip\hfil\cr
        TAUP--2633--00 \cr
        UTTG--13--00 \cr
        }}}
\vskip 1cm

\bigskip
\centerline{\LARGE\bf Domain Walls in the Large $N$ Limit%
    \footnote{Research supported in part by
        the US--Israeli Binational Science Foundation,
        the Israeli Science Foundation,
        the German--Israeli Foundation for Scientific Research (GIF),
        the US National Science Foundation
        (grants PHY--95--11632 and PHY--00--71512)
        and the Robert A.~Welsh Foundation.
        }}
\bigskip\bigskip
\centerline{\bf Yuval Artstein$\strut^{(a)}$,
        Vadim S.~Kaplunovsky$\strut^{(b)}$
        and Jacob Sonnenschein$\strut^{(a)}$}
\medskip
\begingroup \leftskip=18mm \rightskip=18mm plus 5cm \it
    \noindent\llap{(a)\enspace}%
    School of Physics and Astronomy\\
    Beverly and Raymond Sackler Faculty of Exact Sciences\\
    Tel Aviv University\\
    Ramat Aviv, Tel Aviv 69978, Israel
    \par\smallskip
    \noindent\llap{(b)\enspace}%
    Theory Group, Physics Department\\
    University of Texas\\
    Austin, TX~78712, USA
    \par\endgroup

\bigskip\bigskip
\begin{abstract}
We study BPS saturated domain walls in supersymmetric $SU(N)$
Yang--Mills theories in the large $N$ limit.
We focus on the Seiberg--Witten regime of ${\cal N}=2$ theory
perturbed by a small mass ($m<O(\Lambda/N^4)$) and
determine  the wall profile by numerically
minimizing its energy density.
Similar to the $SU(2)$ wall studied in
a previous work, the $SU(N)$ wall has a five layer structure, with two
confinement phases on the outside, a Coulomb phase in the middle
and two transition regions.
\end{abstract}
\end{titlepage}


\section{Introduction}

The phenomena of
 Domain walls, namely,  classical configurations that
  interpolate  between degenerate
discrete vacua arise in many areas of physics.
In fact even in high energy physics  domain walls show up in very distinct
scenarios   like in supergravity,   in grand unification
and in supersymmetric gauge
theories.
In this note we focus on the latter framework.
Domain walls in this context
have been a subject of a rather intensive study in recent
years \cite{KovShi} -\cite{ gs:00}.
Theories  with   spontaneously broken discrete symmetries are
natural setups  for domain walls. A well known  prototype of such
a theory is the
 $\NS=1$
Supersymmetric $SU(N)$ Yang--Mills theory.
It  has a non-anomalous $Z_{2N}$
chiral symmetry which is spontaneously broken down to the $Z_{2}$
group by the
expectation value of the gaugino bilinear
$\left\langle \mathop{\rm tr}\lambda^\alpha\lambda_\alpha\right\rangle$.
 The SYM theory thus has $\N$
degenerate discrete vacua, each characterized by a different value of the
chiral gaugino condensate.

In supersymmetric theories there is a special class of domain walls,
the so-called BPS-saturated domain walls which
preserve half of the  supersymmetries of the theory.
In terms of the field configuration of the domain wall,
preserving supersymmetries imposes
first-order differential equations of the general form
\be\label{BPS}
(Q_{\alpha}-ie^{i\varphi}\sigma^3_{\alpha\dot\alpha}\bar Q^{\dot\alpha})
\left|\hbox{\rm wall}\right\rangle =0,
\ee
where the wall is in the $x_1,x_2$ plane with the fields varying in
the normal direction $x_3$, $Q_\alpha$ are the supersymmetry generators,
and $\varphi=\arg(\Delta W)$ is the phase of the superpotential
difference  $\Delta W$ between the two vacua
separated by the wall.
The solutions to these BPS equations automatically solve
the second order field equations of motion.

 Like other BPS-saturated states, the BPS domain walls are more
tractable and one may reasonably hope for some exact results for such
walls even in a context of a confining strongly interacting theory.
Indeed, the tension i.e. energy per unit area of a BPS domain wall is
exactly determined by the difference between the superpotential values in
the two vacua connected by the wall
without having to find the actual wall profile,
\be
 \epsilon  \equiv\ {\rm Energy\over
Area}\ =2|\Delta W|
\label{epsilon}
\ee
 In the $\NS=1$ SYM theory, the
superpotential --- which acts as a central charge for domain walls --- is
related by the chiral anomaly to the gaugino condensate, so a BPS domain
wall has tension\cite{DvaShi}
$\epsilon = {\N\over 8\pi^2}\, \left|\Delta
    \left\langle\mathop{\rm tr}\lambda^\alpha\lambda_\alpha\right\rangle
    \right| $.

There are two basic ingredients  that determine the  BPS equation (\ref{BPS}),
the superpotential  and the  K\"ahler  potential.
 Whereas
 the  exact form of the effective superpotential
for  the $\NS=1$ SYM theory  is known,
 the  K\"ahler  potential is not fully determined. The effective superpotential
is  constrained by the
requirements of  holomorphy and flavor symmetry
\cite{VenYan,VenYanTay,AffDinSei}.
Unfortunately, no such constraints apply to the effective K\"ahler
function of the theory which controls the kinetic energies of the fields.
Prior to our previous paper\cite{ksy:99},
 all the investigations of this
issue have {\sl assumed} specific K\"ahler functions, only to find that
the answer depends on their assumptions.
In fact, the singularities of a particular
K\"ahler metric led the authors of
\cite{KovShi,DvaShi,ks:97} to claim that  the SYM theory has an additional
chiral-invariant vacuum --- despite overwhelming evidence that it does not.
%
%

Such  problems of determining the BPS equations can be avoided
in theories with more supersymmetries.
 Indeed, the
situation is under much better control for the $\NS=2$ SQCD where the
K\"ahler metric follows from a holomorphic pre-potential and the entire
low-energy effective Lagrangian is completely determined by the
Seiberg--Witten theory \cite{sw:94}.  In the $\NS=1$ terms, the $\NS=2$ SQCD
has an extra chiral superfield in the adjoint representation of the gauge
group. Giving this superfield a mass $m\neq0$ breaks the supersymmetry
down to $\NS=1$. In the $m\to\infty$ limit, the adjoint superfield
decouples from the low-energy physics and one is left with an effective
$\NS=1$ SQCD. We have therefore decided to study the BPS-saturated domain
walls in the $\NS=2$ SQCD perturbed by the adjoint mass $m$.

The analysis of the  the  BPS domain walls  in the
 $SU(2)$ Seiberg--Witten (SW) theory perturbed by  a small
adjoint mass, $m\ll\Lambda$, was performed in \cite{ksy:99}.
In the small mass regime,
it was found that the  SW domain wall has a sandwich-like
five-layer structure.
In each of the two outer layers, the fields asymptote to their respective
vacuum values.  This behavior corresponds to the two confining phases of the SW
theory characterized by respectively magnetic monopole or dyon
condensates. In the middle layer,  the theory  is in its Coulomb phase,
 and the
modulus field slowly interpolates between its stable-vacuum values; for
$\rm mass\neq0$, the Coulomb phase is thermodynamically unstable in bulk
but exists in a layer of finite thickness inside the domain wall. The two
remaining layers contain transition regions between the Coulomb and the
appropriate confining phases.
This profile was determined  by solving numerically the BPS equations
for each of the wall's layers.
 It was found  that the mass is restricted to the region
 $m\leq\Lambda/400$.  Beyond this limit, the transition regions
take over the Coulomb phase region and overlap each other. Also, the wall
becomes too thin to be analyzed in terms of a low-energy i.e
long-distance effective theory such as Seiberg--Witten;
Nevertheless, it was  argued that the
BPS-saturated domain wall exists for any $m$, small or large.
In section 2 we summarize the analysis and results of \cite{ksy:99}.

\par\indent\hskip 0pt minus 3pt
The original $SU(2)$ Seiberg--Witten theory was  extended \cite{af:95,klyt:95}
to  $SU(N)$ gauge groups. Since these theories are more complicated
than  the $SU(2)$ theory, so is the analysis of their  domain wall
configurations.
However, one may anticipate simplifications by using  the large $N$ limit.
Indeed such simplifications were found in \cite{ds:95} where the large $N$
limit of the $SU(N)$ SW theories were addressed.
In general the $SU(N)$  ${\cal N}=2$ SW theory has
a moduli space of $N-1$ complex dimensions.
There are $N$ points at which $N-1$ monopoles become massless, which become the
vacuum points of the theory when perturbed by a superpotential
breaking the number of supersymmetries down to one.

Douglas and Shenker \cite{ds:95} introduced a
trajectory  along the $N=2$ moduli space  that  connects the
extreme semiclassical regime  to  the singular point.
They showed that along this trajectory the one
loop expression for $a_D$ is exact apart from a vanishingly small region around
 the singularity.
 To analyze the domain wall profiles we introduce a trajectory that
interpolates between adjacent vacua by
analytically continuing the trajectory of  \cite{ds:95}.

The object of this work is to solve the equations  and extract the profiles
of the  BPS saturated domain-walls  in the SW
theory with $SU(N)$ gauge group in the large $N$ limit,
perturbed by a mass term.
The equations for a BPS wall appear to be too complicated to solve directly.
Instead the wall is constructed in two steps:
firstly, by regarding only a subset
of the BPS equations and disregarding the rest, a first approximation of a
domain wall profile was calculated.
Then by a numerical method the configuration was iteratively
deformed into a state of minimal energy.
This two step procedure  is described is section 3.
We found that  the final
wall configuration is again  (i) limited to the small values of
$m$ and (ii) characterized by a five-layer profile.
In section 4 we summarize the results and discuss some open questions.

\section{The SU(2) domain wall}

Consider the Seiberg--Witten theory with gauge group $SU(2)$.
When perturbed by a superpotential breaking supersymmetry down from ${\cal
N}=1$ to ${\cal N}=2$,
the moduli space of vacua collapses into two distinct vacuum points. At one of
these vacua, a massless monopole
field condenses, and at the other a dyon field condenses. The low energy
effective action is thus a different
theory in each of these vacua, and when a domain wall stretches between them,
different theories rule
its different regions. The different theories cannot coexist in the same region
of space,
and must be spatially separated in order for the
configuration to make sense. Along the domain wall, three distinct phases can
be identified: an electric-confinement
phase where monopoles exist, an oblique-confinement phase with dyons, and a
Coulomb phase, away from the two vacua
where neither monopoles nor dyons are allowed.

The expectation value of $u=\mathop{\rm tr}(\phi^2)$ serves as the global
coordinate
of the ${\cal N}=2$ moduli space, for the construction of the domain wall.
The perturbation term for the superpotential is
$m\cdot u$. The K\"ahler metric for this coordinate, $g_{u\bar{u}}$,
is found by numerically
calculating the Seiberg--Witten elliptic integrals \cite{sw:94}. The
resulting metric is a function of $u$ in the domain
$[-\lambda,\lambda]$, which diverges logarithmically
towards both ends, and is rather flat otherwise. It was found in \cite{ksy:99}
to be quite accurately approximated by the function
\begin{equation}
g_{u\bar{u}}(u) \approx \frac{1}{16\pi^2 \Lambda^2}
\log\frac{64\Lambda^4}{\Lambda^4-u^2}.
\label{metric}
\end{equation}
The metric for the monopoles is taken canonically to be
$g_{M\bar{M}}=g_{\tilde{M}\bar{\tilde{M}}}=1$ and similarly
for the dyon fields.
For the superpotential, one must look separately at different regions of $u$,
and determine it separately for
the three different phases of the theory. The perturbed superpotential used for
the electric-confinement phase is
\begin{equation}
W=m\cdot u+\frac{\Lambda^2-u}{i\sqrt{2}\Lambda}M\tilde{M}.
\label{su-pot}
\end{equation}
At the oblique-confinement region a similar superpotential can be written for
the massless dyons:
\begin{equation}
W=m\cdot u + \frac{\Lambda^2+u^2}{-i\sqrt{2}}\Lambda D\tilde{D}.
\end{equation}
Away from the vacuum points, in the Coulomb phase, the only term in the
superpotential is $m\cdot u$.
This makes sense if the expectation value of the monopoles
vanishes as we get away from $u=-\Lambda$, and that of the dyons vanishes away
from $u=\Lambda$.
In the construction to follow this condition is met, and imposes a limitation
on the value of the mass~$m$.

Generally,
the field profile of a BPS-saturated domain wall is governed by
eqs.~(\ref{BPS}).
In an effective field theory without higher-derivative terms in its
Lagrangian, the explicit form of these equations is
\begin{equation}
\frac{dA^i}{dx_3}=e^{i\varphi}g^{i\bar\jmath} \frac{\partial
W^*}{\partial\bar{A}^{\bar\jmath}}
\label{bps}
\end{equation}
where $A^i(x_3)$ runs over all the scalar fields of the theory.
For the theory at hand, this means
$u$, $M$, $\tilde M$, $D$ and $\tilde D$.
Fortunately, in the Seiberg--Witten regime of the theory (small $m$),
the monopole and the dyon fields are segregated to different regions of space,
so for any particular stretch of $x_3$ we may eliminate some of the fields
{}from the BPS eqs.~(\ref{bps}).

Altogether, the wall has five distinct layers:
On the left side, the $u$, $M$ and $\tilde M$ fields asymptote to their vacuum
values
in the electric confinement phase and there are no dyons.
Next, there is a transition region where monopole fields turn off.
The third, middle layer is in the Coulomb
phase, where both monopoles and dyons are absent and $u$ varies slowly from
$\Lambda^2$ to $-\Lambda^2$.
Next, there is another transition layer where the dyon fields turn on.
Finally, on the right side, $u$, $D$ and $\tilde D$ fields approach
their vacuum expectation values in the
 oblique confinement phase.

\begin{figure}[t]
\begin{center}
\psfig{figure=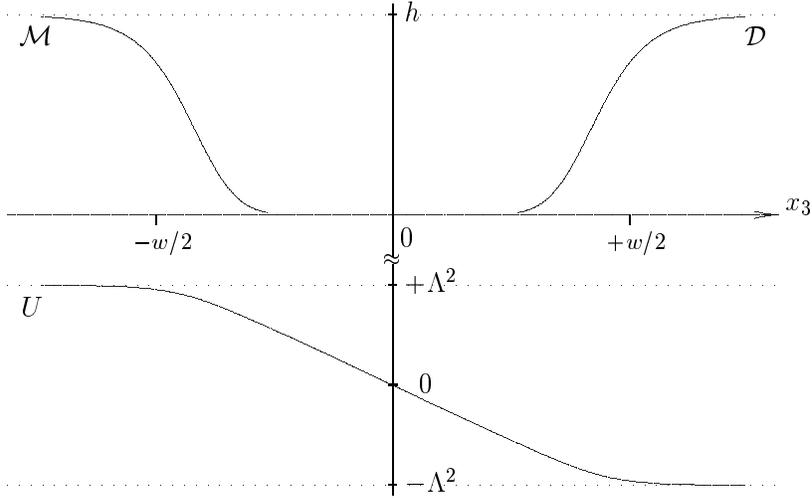,width=15cm}
\end{center}
\caption{\sl
    Field profiles for the Seiberg--Witten domain wall, $(m=\Lambda/2000)$.
    \rm (Figure 5 of \cite{ksy:99})}
\label{f5}
\end{figure}

Consider the middle layer first.
The only important field here is $u$, hence the BPS eqs.\ reduce to
\begin{equation}
\frac{du}{dx_3} = m^*e^{i\varphi}g^{u\bar{u}}.
\end{equation}
Without loss of generality, we assume real positive $m$ and $\Lambda$, hence
$e^{i\varphi}=-1$ and real $u(x_3)$.
For the approximate metric~(\ref{metric}), the exact solution is
\begin{equation}
16\pi^2mx_3\ =\ \log\frac{\Lambda^2+u}{\Lambda^2-u}\
-\ \frac{u}{\Lambda^2}\log\frac{64e^2\Lambda^4}{\Lambda^4-u^2}
\end{equation}
and the domain-wall has finite thickness (as defined in \cite{ksy:99})
of about $w=0.0605m^{-1}$. Working numerically with the exact metric yields a
slight adjustment to this value, $w=0.0625m^{-1}$.

The two transition regions are isomorphic, so it is enough to consider the
transition between the electric-confinement
and the Coulomb phases.
The relevant fields here are $u,M$ and $\tilde{M}$, the superpotential is
(\ref{su-pot}) and thus the  BPS equations are
\begin{eqnarray}
\label{uMM}
g_{u\bar{u}}\frac{du}{dx_3} & = & -m+\frac{1}{\sqrt{2}\Lambda}M^*\tilde{M}^*,
        \nonumber\\
\frac{dM}{dx_3} & = & \frac{\Lambda^2-u^*}{i\sqrt{2}\Lambda}\tilde{M}^*,\\
\frac{d\tilde{M}}{dx_3} & = & \frac{\Lambda^2-u^*}{i\sqrt{2}\Lambda}M^*.
        \nonumber
\end{eqnarray}
As in the Coulomb phase, the $u$ field takes real value while
the magnetic gauge symmetry allows us to specify real $M$ and imaginary
$\tilde M=iM$.
The boundary conditions for eqs.~(\ref{uMM})
amount to requirements that 1.
all fields must asymptotically approach their vacuum expectation
values for $x\to{-\infty}$, and 2. the monopole fields $M,\tilde{M}$ must
disappear towards the middle of the wall so that the solution can be matched
with the previous solution of the Coulomb phase.
Of course, analytically the monopole fields cannot vanish exactly,
but numerically they do become negligibly small at $x_3=0$
--- drop to less than 1\% of their vacuum values ---
provided $m$ is smaller than about $\Lambda/400$.

The numerical wall profile for $m=\Lambda/2000$ is shown in figure~(\ref{f5});
note the transition regions occupying a considerable fraction
of the wall width despite rather small $m$, although
this fraction becomes smaller for even smaller values of the perturbing mass
parameter.
On the other hand, for larger perturbing masses, the transition regions spread
towards the mid-wall point and start overlapping each other.
Both the monopole and the dyon fields appear to be present in such an overlap,
a situation quite untractable in terms of an effective field theory.
In fact, this is just a symptom of a more general problem:
A larger $m$ makes for a thinner wall, which can never be adequately described
in terms of a low-energy --- {\it i.e., \sl long-distance} --- effective
theory theory such as Seiberg--Witten.

\section{The large-$N$ domain wall}

The ${\cal N}\approx2$ limit of the $SU(N)$ SYM is generally similar
to the Seiberg--Witten $N=2$ theory, but instead of a single complex
modulus $u=\mathop{\rm tr}(\phi^2)$ we now have $N-1$ moduli.
The eigenvalues $\phi^1,\ldots,\phi^N$ of the scalar field $\phi$
(viewed as an $N\times N$ matrix) provide a redundant coordinate system
for this moduli space (note $\sum_j\phi^j=0$).
At $N$ discrete points in this space~\cite{ds:95},
\begin{equation}
\label{Stable}
\left\langle\phi^j\right\rangle\
=\ e^{i\pi r/N}\times 2\Lambda\cos{\pi(j-1/2)\over N}\,,\qquad
r=0,1,\ldots,(N-1),
\end{equation}
the maximal number of magnetic monopoles or dyons become massless.
Such massless monopoles/dyons acquire non-zero vacuum expectation values
when the ${\cal N}=2$ theory is perturbed by the superpotential
\be
\label{Wpert}
W\ =\ {Nm\over2}\,\sum_j(\phi^j)^2\,,
\ee
(the ${\cal N}=1$ mass term) and the theory enters a confining phase
(electric or oblique).
All vacua other than~(\ref{Stable}) become unstable.

Following M.~Douglas and S.~Shenker~\cite{ds:95}, let us Fourier transform
the eigenvalues $\phi^j$ according to
\begin{eqnarray}
t^k & = & \frac{2}{N} \sum_{j=1}^N \phi^j \cos{\pi k(j-1/2)\over N}\,, \\
\phi^j & = & \sum_{k=1}^{N-1} t^k \cos{\pi k(j-1/2)\over N}\,.
\end{eqnarray}
In terms of the new moduli $t^1,\ldots,t^{N-1}$, the $N$ stable vacua
are located at $t^1=2\Lambda e^{\pi i r/N}$ and $t^2=t^3=\cdots=t^k=0$.
Let us focus on the $r=0$ vacuum (electric confinement) where the light
fields comprise the $t^k$ moduli and the magnetic monopoles
$M_l,\tilde M_l$ ($l=1,\ldots,N-1$).
The superpotential for these fields combines the perturbative
term~(\ref{Wpert})
and the monopole mass terms $\sum a^D_l(t)M_l\tilde M_l$.
Expanding the latter around $t^k\approx 2\Lambda\delta_{k,1}$ and taking the
large $N$ limit~\cite{ds:95}, we arrive at
\begin{equation}
W\ =\ \frac{N^2m}{4}\sum_k (t^k)^2\
+\sum_{k,l}\sin\frac{kl\pi}{N}(t^k-2\Lambda\delta_{k,1})M_l\tilde{M}_l.
\label{W}
\end{equation}
Among other things, this superpotential governs the monopole condensations;
solving for $dW=0$, we find
\begin{equation}
\left\langle M_l\tilde{M}_l\right\rangle\ =\ -2N\Lambda m\sin\frac{\pi l}{N}\,.
\label{monopoles}
\end{equation}

\subsection{Domain Wall: First Approximation}

We would like to construct a BPS-saturated domain wall between two adjacent
vacua~(\ref{Stable}), say $r=0$ and $r=1$.
Both vacua have $t^2=\cdots=t^{N-1}=0$,
so as a first approximation to the domain wall's profile, we assume
all the $t^{k>1}$ moduli to maintain zero values throughout the wall while
the $t^1$ modulus evolves from $t^1=2$ to $t^1=2e^{\pi i/N}$.
(Henceforth we shall use $\Lambda=1$ units).
This `trajectory' through the moduli space is related to the `scaling
trajectory'
\be
\phi^j(s)=e^{s/N}\times 2\cos\frac{\pi(j-1/2)}{N}
\ee
of ref.~\cite{ds:95} via analytic continuation to imaginary $s$ that runs
{}from 0 to $\pi i$ ($t^1=2e^{s/N}$).
Along this trajectory, the moduli metric is diagonal in the $t^k$ basis
(in the large $N$ limit only!),
\be
g_{\bar k k'}\ =\ \delta_{\bar k,k'}\times
{\mathop{\rm Re}(FG^*)\over 8\pi\sin{\pi k\over 2N}}\,,
\label{kmetric}
\ee
where $F(s,\frac{k}{N})$ and $G(s,\frac{k}{N})$ are defined in eqs.~(5,9--10)
of~\cite{ds:95};
for $k\ll N$, $F\approx G\approx 1$ regardless of $s$.
For larger $k=O(N)$, $F$ and $G$ --- and hence the metric $g_{\bar kk}$ ---
become $s$ dependent, but fortunately, the detailed nature of this dependence
is not germane
for the present analysis.

Similarly to the $SU(2)$ case, we expect the domain wall to have the five-layer
structure with two well-separated transition regions for sufficiently small
perturbing mass~$m$.
Consequently, there should be no monopole fields for $x_3\ge0$ or dyon
fields for $x_3\le 0$.
Physically, the wall is left-right symmetric, so let us focus on its left half
and thus dispense with the dyon fields.
The remaining moduli and monopole fields are governed by the BPS equations
\begin{eqnarray}
g_{k\bar k}\frac{\partial t^{*\bar k}}{\partial x_3} & = &
    e^{i\varphi}\, \frac{\partial W}{\partial t^k}\,, \label{b1} \\
\frac{\partial M_l^*}{\partial x_3} & = &
    e^{i\varphi}\, \frac{\partial W}{\partial\tilde{M}_l}\,, \label{b2} \\
\frac{\partial\tilde{M_l}^*}{\partial x_3} & = &
    e^{i\varphi}\, \frac{\partial W}{\partial M_l}\,. \label{b3}
\end{eqnarray}
In the Coulomb phase domain where the monopole fields have very low values,
the right hand sides of these equations vanish, except for the the first
eq.~(\ref{b1})
for the $t^1$ modulus.
Thanks to the diagonality of the moduli metric~(\ref{kmetric}), this means that
$t^1$ is the only non-constant field in the Coulomb domain, in perfect
consistency with the $t^{k>1}\equiv0$ trajectory assumption.
Unfortunately, in the middle of the transition region where the monopole fields
neither vanish nor take vacuum values, the right hand sides of eqs.~(\ref{b1})
for the moduli $t^2,t^3,\ldots,t^{N-1}$ do not vanish, which means that the
$t^{k>1}\equiv0$ wall is not exactly BPS.

Nevertheless, as our first approximation, we would like to freeze all
$t^{k>1}\equiv0$
and solve the BPS equations for the $t^1(x_3)$ and all the monopoles $M_l(x_3)$
and
$\tilde M_l(x_3)$ analytically.
In the following subsection, we numerically calculate the deviation of the true
BPS
wall from this trajectory; we shall see that such deviations are negligible in
the
Coulomb domain but significant in the transition region, especially its outer
edge.
For the overall profile of the wall however, $t^{k>1}\approx0$ is a good
approximation.

For the two vacua $r=0$ and $r=1$, $\Delta W=N^2m(e^{2\pi i}-1)\approx 2\pi Nm
i$,
so let us re-phase $W\to -iW$ and make the eqs.\ (\ref{b1})[for the $T^1$]
and (\ref{b2}--\ref{b3}) real.
The boundary conditions are $t^1=2$ and eqs.~(\ref{monopoles}) for the monopole
fields
at $x_3=-\infty$, while for $x_3=0$,
$t^1=2e^{\pi i/2N}\approx 2+\pi i/N$ and the monopoles must vanish.
The BPS solution can be generally written as
\begin{eqnarray}
M_l=\tilde{M}_l & = & i\sqrt{2Nm\sin(\frac{\pi l}{N})}\times
    \exp\Bigl( -h\sin\frac{\pi l}{N}\Bigr),
    \label{mono}  \\
t^1 & = & 2+\frac{2is^1}{N}\,, \label{sdef}
\end{eqnarray}
with $s^1(x_3)$ and $h(x_3)$ developing according to differential equations
\begin{eqnarray}
\frac{\partial s^1}{\partial x_3}&=&
    2\pi^2Nm[N-2\sum_l\sin^2(\frac{\pi l}{N})e^{-2h\sin\frac{\pi l}{N}}] ,
    \label{ds} \\
\frac{\partial h}{\partial x_3} & = & \frac{2s^1}{N} \,.
    \label{dh}
\end{eqnarray}
Physically, $s^1$ simply keeps track of the coordinate $t^1$ of the ${\cal
N}=2$
moduli space, while $h$ governs the decay of the monopole condensates away from
the
confining vacuum domain;
note different monopole condensates $M_l$ decaying at different rates
thanks to the $\sin(\frac{\pi l}{N})$ factor in the exponent in
eq.~(\ref{mono}).

The boundary conditions are $s^1\rightarrow 0$ and $h\rightarrow 0$ at
$-\infty$, and $s^1=\pi/2$ at $x_3=0$.
Also, {\em all} the monopole condensates must become vanishingly small towards
$x_3=0$,
thus $h(\pi/N)\gg 1$ at $x_3=0$.
Physically, the last requirement imposes a lower limit on the domain wall's
width
and hence an upper limit on the perturbing mass~$m$.
Indeed, in light of eq.~(\ref{dh}) we estimate $h(x_3=0)\sim
\frac{\pi}{2N}\times\frac{w}{2}$,
hence the wall should be much wide than $w_0\sim 4N^2/\pi^2$.
On the other hand, eq.~(\ref{ds}) lets us estimate $2\pi^2
N^2m\times\frac{w}{2}\sim\frac\pi2$,
hence
\be
w\sim \frac{1}{2\pi N^2 m}\,. \nonumber
\ee
Consequently, to maintain spatial separation between the monopole and the dyon
fields,
we need
\begin{equation}
m\ll \frac{\pi}{8N^4}
\end{equation}
(in $\Lambda=1$ units).

\begin{figure}[p]
\begin{center}
\psfig{figure=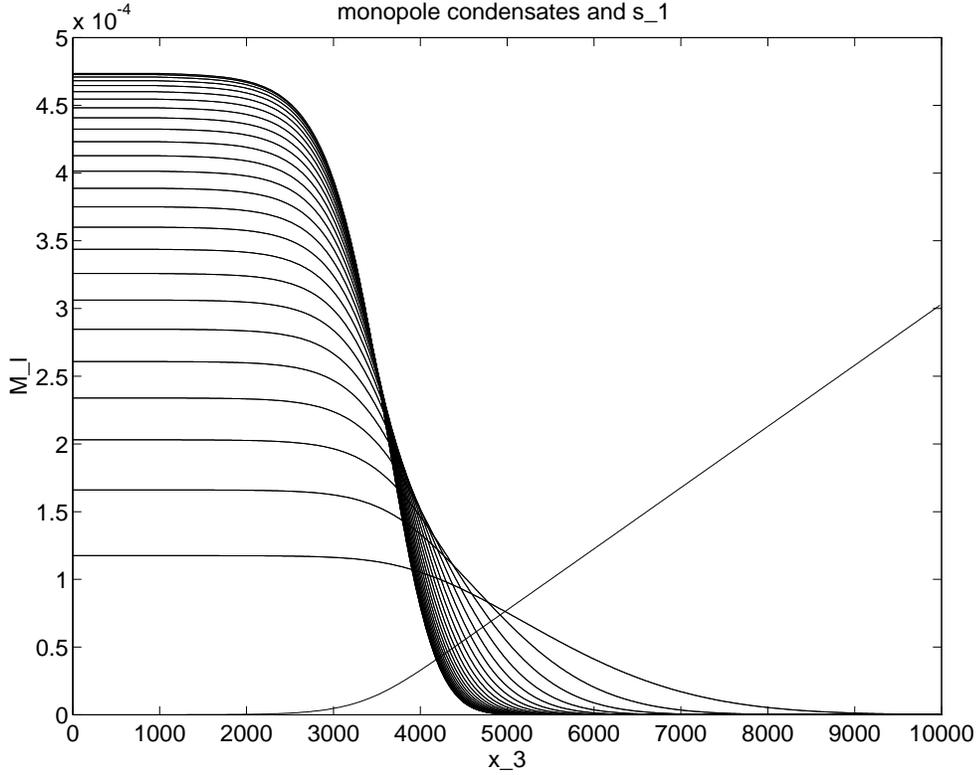,width=15cm}
\end{center}
\caption{%
    First approximation to the transition region of the domain wall,
    between the electric confinement phase on the left and the Coulomb phase on
the right.
    One can see how the monopole condensates disappear towards the middle of
the wall
    (the right side of the picture), while the $s^1$ coordinate moves from 0 up
to $\frac\pi2$
    at the center (far right, out of the frame).
    The horizontal axis is $x_3$ in units of $1/ \Lambda$, shifted so that the
zero point
    is deep in the confining phase on the left.
    The scale for the monopole condensates is shown on the left;
    the scale of $s^1$ is 2500 times large, thus $s^1=0.75$ at the right edge
of the frame.
    }
\label{pr11}
\end{figure}

Given small enough $m$, equations (\ref{ds}--\ref{dh}) can be easily solved
numerically.
Figure~(\ref{pr11}) displays the resulting field profiles through the
transition region
of the wall, as calculated for $N=51$, $m=2.2\cdot 10^{-9}$.
Note different monopole condensates decreasing at different rates but all
reaching
negligible values in the Coulomb domain (the right side of the figure).

\subsection{Correcting the Trajectory:  Minimizing the Energy}

The net energy density of a  BPS-saturated domain wall
comprises potential and  kinetic terms, each contributing $|\Delta W|$ to the
total.
The approximately but not exactly BPS domain wall of the previous section
has correct kinetic energy,
\begin{eqnarray*}
g_{i\bar{j}}\partial_3A^i\partial_3\bar{A}^{\bar\jmath}\ & =\ &
g_{11}\partial_3t^1\partial_3t^{*1}\ +
     \sum_l\left(\partial_3M_l\partial_3{M}_l^*
         + \partial_3\tilde{M}_l\partial_3\tilde{M}_l^*\right) \\
& =\ &
\partial_3t^1\frac{\partial W}{\partial t^1}\
    + \sum_l\left(\partial_3M_l\frac{\partial W}{\partial M_l} +
            \partial_3\tilde{M}_l\frac{\partial W}{\partial\tilde{M}_l}\right)
\\
& =\ & \partial_3 W ,
\end{eqnarray*}
thus
\be
\epsilon_{kin}\ =\int_{-\infty}^\infty\!\!dx_3\, \partial_3 W\ =\ \Delta W.
\ee
Its potential energy however is higher than BPS because of un-satisfied BPS
equations
(\ref{b1}) for the $t^{k>1}$.
Indeed,
\begin{eqnarray}
\epsilon_{pot}\ &= &
\int_{-\infty}^{\infty}\!\!dx_3\, g^{i\bar\jmath}\frac{\partial W}{\partial
A_i}
    \frac{\partial W^*}{\partial\bar{A_{\bar\jmath}}} \nonumber \\
&=& \int\!\!dx_3\,\biggl[
        \sum_k g^{k\bar{k}}\frac{\partial W}{\partial t^k}\frac{\partial
W^*}{\partial t^{*k}}\
        + \sum_l\left(\frac{\partial W}{\partial M_l}\frac{\partial
W^*}{\partial M_l^*}\,
                +\,\frac{\partial W}{\partial\tilde{M}_l}
                    \frac{\partial W^*}{\partial\tilde{M}_l^*})\right)
        \biggr]  \nonumber \\
&=& \int\!\!dx_3\,\biggl[\partial_3 t^1\,\frac{\partial W}{\partial t^1}\,
        +\sum_l\left((\partial_3 M_l\frac{\partial W}{\partial M_l}
                +\partial_3\tilde{M_l}\frac{\partial
W}{\partial\tilde{M}_l}\right)\,
        +\sum_{k\neq 1}g^{k\bar{k}}\left|\frac{\partial W}{\partial
t^k}\right|^2
        \biggr] \nonumber \\
&=& {}\ \Delta W\
        +\int\!\!dx_3\sum_{k\neq 1}g^{k\bar{k}}\left|\frac{\partial W}{\partial
t^k}\right|^2.
\label{e-pot}
\end{eqnarray}
Altogether, the last term in eq.~(\ref{e-pot}) is the excess energy density
over the BPS
domain wall due to violation of some of the BPS equations.
Spacewise, this excess energy comes solely from the transition region
($\partial W/\partial t^k=0$ in both Confining and Coulomb phases),
thus our approximate solution is accurate for the Coulomb phase region ---
which
constitutes most of the wall width and provides most of its energy in the
small~$m$ limit.
However, the transition regions of the wall are rather interesting even when
they are narrow, so we would like a better approximation to the field profiles.
In particular, we need to get rid of the wrong assumption about $t^{k>1}$ and
calculate
their actual profiles.

To obtain such a better approximation, we minimize the wall's net energy
density as a function of its field profile.
Basically, we first discretize $x_3$ and analytically evaluate the variational
derivatives
of the energy with respect to all possible lattice field variations, then
numerically
evolve the field configuration in the direction of the steepest energy descent.
The boundary values are fixed to vacuum values at the left boundary and
all fields except $t^1$ are zero at the right boundary.
We start with the approximate wall profile obtained in the previous subsection,
then
evolve the lattice fields until we reach the (numeric) minimum of the wall's
energy.

Our calculations are somewhat eased by the
symmetries of the superpotential~(\ref{W}) and hence of the exact solution
to the BPS equations,
\be
M_l\equiv\tilde M_l\equiv M_{N-l}\equiv\tilde M_{N-l}\quad
{\rm and}\quad t^k\equiv0\ \forall\,{\bf even}\ k .
\label{symmetries}
\ee
Naturally, we hardcoded these symmetries into our minimization procedure.
Also, for convenience of taking the large~$N$ limit, we use rescaled moduli
fields
\begin{equation}
s^k=\frac{N}{2i}(t^k-2\delta_{k,1}) \,;
\label{s^k}
\end{equation}
note consistency with the $s^1$ variable used in eq.~(\ref{sdef}).
In terms of the independent $M_l(x_3)$ and $s^k(x_3)$,
the energy density we need to minimize is
\be
\epsilon\ =\int\!\!dx_3\rho \nonumber
\ee
where
\begin{eqnarray}
\rho(x_3)& = &
\sum_{{\rm odd}\,k}\left\{ \frac{N^2 g_{\bar kk}}{4}\,(\partial_3 s^k)^2\
        +\ \frac{4}{g_{\bar kk}}\left(
                \smash{\sum_{l=1}^{(N-1)/2}} M_l^2\,\sin\frac{\pi kl}{N}\
                -\ \frac{N^2m}{2}\,\delta_{k,1} \right)^2
        \right\} \! \nonumber \\
&{}&
+\sum_{l=1}^{(N-1)/2}\left\{ 4\bigl(\partial_3 M_l\bigr)^2\
        +\ \frac{16}{N^2}\,M_l^2\left(
                \smash{\sum_{{\rm odd}\,k}} s^k\,\sin\frac{\pi kl}{N}\right)^2
        \right\}
\label{NetEnergy}
\end{eqnarray}
and
\be
g_{\bar kk}\ \approx\ {1\over 8\pi \sin{\pi k\over2N}}
\label{Gapprox}
\ee
since $\mathop{\rm Re}(FG^*)\approx1$ in eq.~(\ref{kmetric}).%
\footnote{%
        To be precise, this approximmation works very well for $k\ll N$
        but for $k=O(N)$ one should in principle use a more accurate
        approximmation
        \be
        \mathop{\rm Re}(FG^*)\approx
        \Bigl(1-{k\over3N}\Bigr)\Bigl(1-{k\over N}\Bigr)
                \Bigl(1-{k^2\over N^2}\Bigr)
        + \frac{1}{\pi}\,\sin{\pi k\over 2N}\times\log{8\over s^1}\,.
        \ee
        Fortunately, numerically this expression is close enough to 1
        in the region of interest to justify
        the approximmate metric~(\ref{Gapprox}).%
        }

\begin{figure}[p]
\psfig{figure=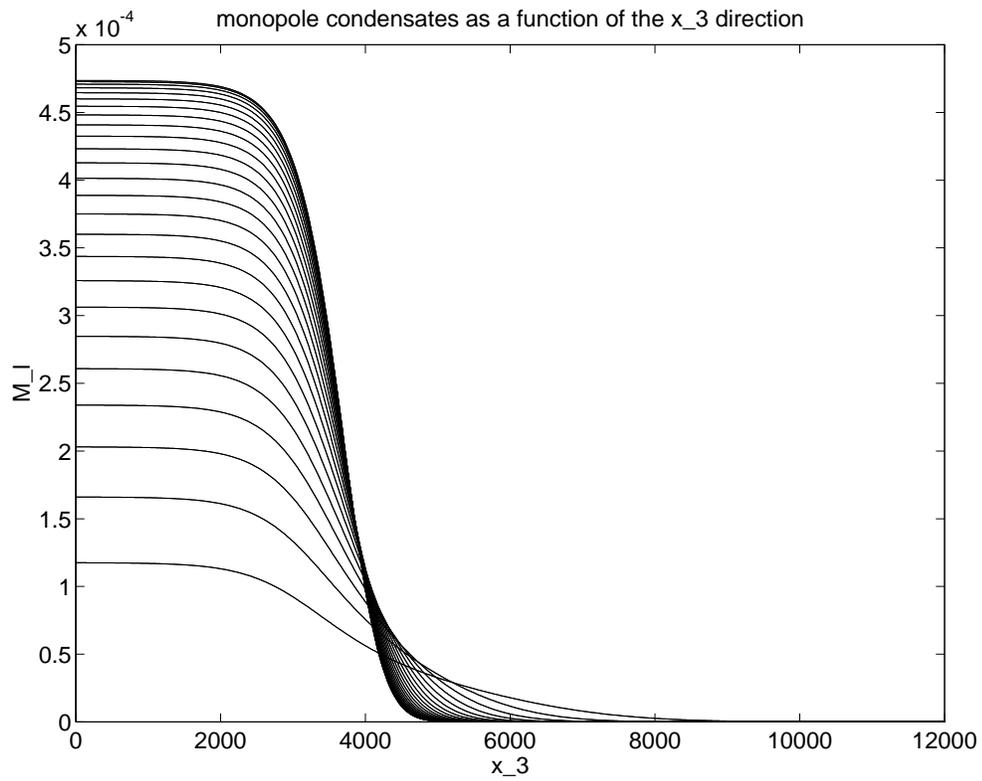,width=15cm}
    \caption{%
    Field profiles for the monopoles at the final domain wall configuration,
    $N=51$ and $m=2.2\cdot 10^{-9}$.
    The highest and steepest line is $M_{25}$, and the lowest $M_1$.
    }
\label{pr12}
\end{figure}

\begin{figure}[p]
\psfig{figure=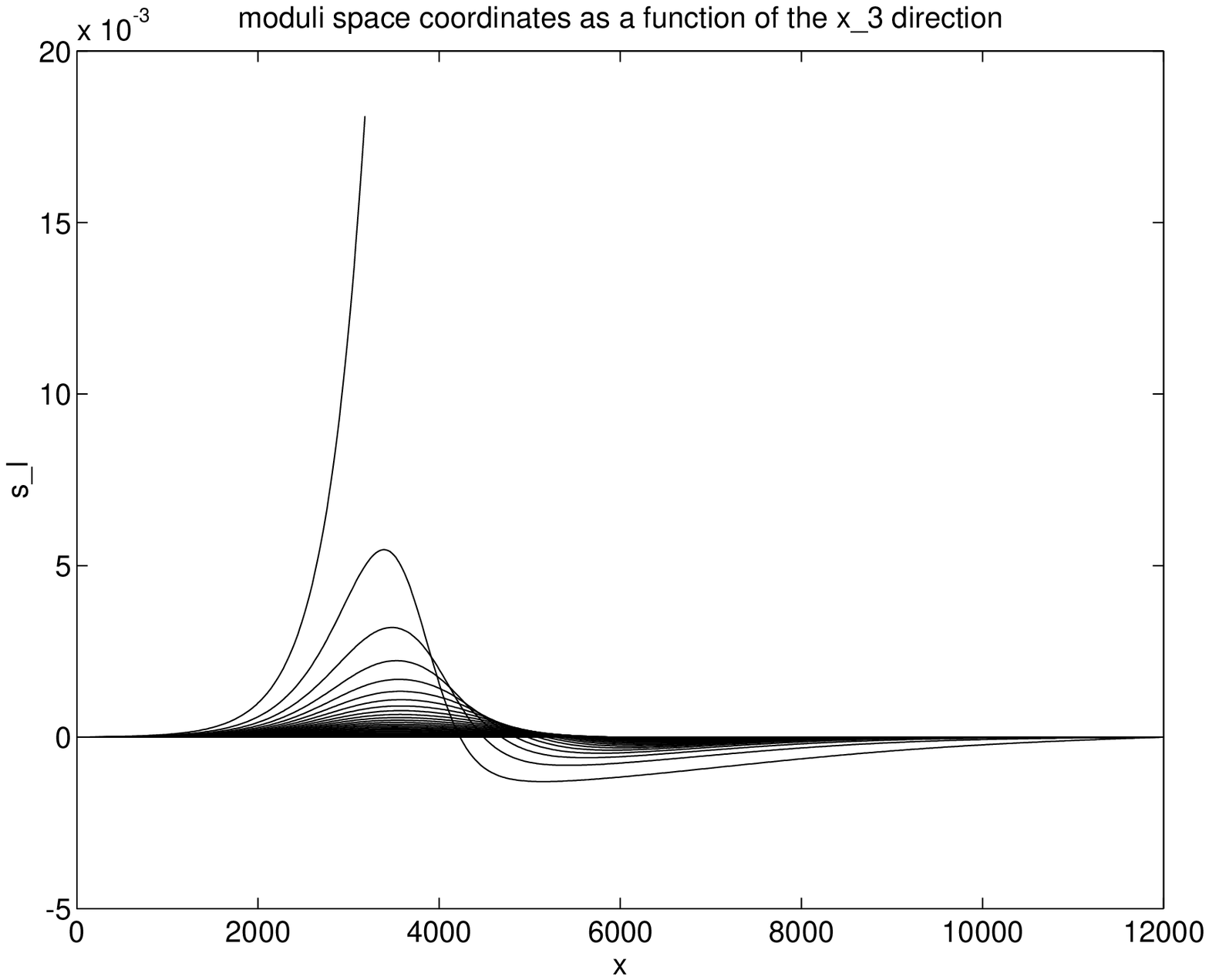,width=15cm}
\caption{%
    Field profiles for the moduli $s^k$ (odd $k$).
    The $s^1$ will continue to rise throughout the Coulomb
    but the plot is cut at $x=3200$ to avoid lines sticking out of the frame.
    The tallest line after $s^1$ is $s^3$, then $s^5$, {\it etc}.}
\label{pr10}
\end{figure}

The technical details of the minimization procedure being rather boring,
let us simply present our results.
Figure~(\ref{pr12}) depicts the monopole condensates profiles --- which have
not
changed dramatically during the energy-minimization procedure, {\it cf.}\
figure~(\ref{pr11}),
except their transition regions became a bit narrower.
This is easily understood in terms of BPS eqs.~(\ref{b1}--\ref{b2}): the
additional terms
involving $t^{k>1}$ moduli make for steeper decrease of the monopole fields.
Figure~(\ref{pr10}) displays the moduli profiles $s^1$ and also $s^3$, $s^5$,
{\it etc.}
As expected, $s^3,s^5,\ldots$ vanish in the Coulomb phase on the right,
but not in the transition region between the confining and the Coulomb phases.
For example, at its peak in the middle of the transition region, the $s^3$
reaches about
18\% of the value of the $s^1$ at the same point.
For other  odd $k$, the $s^k$ reach their peaks at about the same point,
reaching a value roughly proportional to $1/k$.
The reason for such behavior is  the superpotential (\ref{W}):
 for large $k$, $\sin(\pi k l/N)$ oscillates faster
between positive and negative values, leaving an average of order $1/k$.


The width of the transition region can now be
measured by looking at either monopole condensates or moduli~$s^k$.
The highest peak of the $s^3$ field --- where its value is above 50\% of its
maximum --- corresponds to the region where the
highest monopole mode $M_{(N-1)/2}$
decreases from 85\% of its vacuum value down to 15\%.
Other monopoles are wider and recede less steeply.
As for the higher moduli fields, the $s^{k>3}$ have wider peaks,
for example twice as wide for $s^{21}$ than for
$s^3$ and slightly more than that for yet larger $k$.
We see no obvious way to
account for this fact, other than to say that it is a numerical result.
In any case, the high modes are not that much wide wider:
merely by a factor of 2 rather than  $N$ or even $\sqrt{N}$.

Interestingly, on the immediate right of the transition region,
the $s^{k>1}$ moduli drop below zero and develop
small negative values   before they finally disappear in the Coulomb region.
The origin of this behavior is unclear, but this is definitely {\em not} an
artefact
of the boundary condition.
This is verified by pushing the right boundary of the lattice region further
right.

Finally, we would like to mention that when all the $s^k$ are very small, they
have
comparable magnitudes, for example $s^3\approx 0.65 s^1$.
In other words, at the left edge of the wall, where the moduli fields just
begin
to shift away from their values in the confining vacuum, they move in a
direction
at rather wide angle to the $s^1$ ``axis'' characteristic of the Coulomb phase
in the middle of the wall.
This rather unexpected geometry of the wall's `trajectory' of the moduli space
indicates that the present analysis does not extrapolate well into the regime
of not-too-small (${\cal N}=2$)--breaking mass $m$ for which
various regions of the domain wall start overlapping each other.

\subsection{Conclusion}

The structure of the domain wall worked out in this work is in many ways
reminiscent of the $SU(2)$ domain wall.
Both exhibit a similar five-layer profile, and both depend strongly on the
smallness of the perturbation mass  parameter $m$.
The devil is in the details and the $SU(N)$ theories exhibit many new
features associated with multiple monopole/dyon condensates and multiple moduli
fields.
There are many more BPS equations now, all entangled with each other, so
calculating
the field profiles of the wall calls for numerical lattice calculations.
Even the BPS route that the moduli fields take on their way from one stable
vacua
to another is highly turns out to be non-trivial.
In the $SU(N)$ model, the route between two vacua in the moduli space
is not trivial, and distinct functions describe the evolutions of different
monopole fields.

The large $N$ approximation helps us to get explicit expressions for the
K\"ahler metric and for the superpotential, but
unfortunately, at the end of the day we are again faced with the same
limitation as in the $SU(2)$ domain wall, namely that the wall is valid only
for
a small perturbation parameter $m$. In fact, the limitation here is more
severe,
as $m$ must be smaller than ${\cal O}(N^{-4})$.
For such small $m$, the five-layer picture is adequate. The work shows that the
dominant field in the Coulomb phase is
$t^1$, and the other fields either asymptote to zero, or remain very small. It
is not possible, however, to disregard the
other $t$'s in the transition region. As $m$ gets larger and the two transition
regions overlap, on top of the
difficulty from having coexisting monopole fields from two different vacua, we
cannot even say along what route
in the moduli space of the ${\cal N}=2$ theory the domain wall lies. The wall
derived here, however, is a legitimate
domain wall, and it may be interesting to study its low energy excitations in
the future.

\vspace{24pt}
{\large \bf Acknowledgement:}
Research supported in part by
the US--Israeli Binational Science Foundation,
the Israeli Science Foundation,
the German--Israeli Foundation for Scientific Research (GIF),
the US National Science Foundation
(grants PHY--95--11632 and PHY--00--71512)
and the Robert A.~Welsh Foundation.

\renewcommand{\baselinestretch}{1}

\end{document}